\documentclass[aps,prb,superscriptaddress,showpacs,twocolumn]{revtex4}
\usepackage{graphicx}
\begin{document}
\title{Vortex dynamics in the presence of a line of submicron holes along a superconducting microbridge}
\author{J. Bentner}
\affiliation{Institut f\"{u}r experimentelle und angewandte Physik,
Universit\"{a}t  Regensburg, D-93025 Regensburg, Germany}
\author{D. Babi\'{c}}
\altaffiliation[Corresponding author ] {}
\email{dbabic@phy.hr}
\affiliation{Department of Physics, Faculty of Science,
University of Zagreb, Bijeni\v{c}ka 32, HR-10000 Zagreb, Croatia}
\author{C. S\"{u}rgers}
\affiliation{Physikalisches Institut, Universit\"{a}t Karlsruhe, D-76128 Karlsruhe, Germany}
\author{C. Strunk}
\affiliation{Institut f\"{u}r experimentelle und angewandte Physik,
Universit\"{a}t  Regensburg, D-93025 Regensburg, Germany}
%
%
%
%
\begin{abstract}
We measured and compared the electric field vs current density
characteristics in the vortex state of two amorphous
Nb$_{0.7}$Ge$_{0.3}$ microbridges, with and without a line of
submicron holes patterned along the sample axis.
The power dissipation in the perforated sample exhibits a crossover,
being reduced at temperatures well below
the superconducting transition temperature $T_c$ and unexpectedly enhanced 
close to $T_c$.
At low temperatures
the holes are efficient artificial pinning centres and 
reduce the average vortex velocity.
We argue that the dissipation enhancement close to $T_c$
is a consequence of a combination of the
weakened pinning by the holes and an inhomogeneous driving-current
distribution in their vicinity, which results in an increased average
vortex velocity as well as in a channeling of the vortex motion
through the holes.
\end{abstract}
%
%
\maketitle


The rapid development of nanostructuring methods in the last decade
has enabled highly precise fabrication of artificial pinning centres
(APCs) for vortices in superconductors. It has already been proved
that point-like inclusions in a superconducting film, such as
perforating holes,\cite{dald,mosc1} structural defects\cite{matsuda}
and magnetic dots,\cite{martin} can pin vortices. A regular
two-dimensional array of APCs stabilises the vortex lattice against
external driving forces, and results also in commensurability effects
which scale with a matching magnetic field $B_M = \phi_0 / S$, where
$S$ is the area of the array unit cell and $\phi_0$ the magnetic flux
quantum.\cite{dald} At a magnetic field $B > B_M$ vortices fill in
the interstitial positions as well, where they are subject to
intrinsic pinning originating from the structural defects. In a
certain $B$ range above $B_M$ they may even form a commensurate
multiple-flux-quanta lattice, as revealed from magnetisation
measurements.\cite{mosc1} A less symmetrical APC landscape, such as a
rectangular array\cite{velez} or a set of parallel continuous lines
of a magnetic material,\cite{jaque} introduces anisotropies to the
vortex transport. So far the effects of APCs have been found to be important at
temperatures $T$ rather close to $T_c$, typically at reduced
temperatures $t=T/T_c
> 0.9$, whereas at lower temperatures the intrinsic pinning
dominated. Moreover, APCs invariably {\it
decreased the dissipation}, apart from in a very recent
report.\cite{jiang}

The simplest APC is a small perforating hole in a superconductor, having
magnetic permeability larger than its diamagnetic surroundings. The
most efficient pinning is obtained if the hole diameter is comparable
to the magnetic-field penetration depth $\lambda$ rather than the
coherence length $\xi$, as one may expect at first sight. This was
predicted by virtue of the Ginzburg-Landau (GL) theory for a single
hole\cite{take} and observed experimentally for
arrays of holes.\cite{mosc2} The relevance of $\lambda$ for the
artificial pinning potential is a consequence of the comparatively
small supercurrent kinetic energy required for preserving the
flux-quantisation condition around a hole not smaller than $\lambda$.

Local interactions of vortices with holes in the presence of a
driving current have not yet been studied in detail experimentally.
These go beyond the cited matching phenomena and are important for
understanding the function of holes in an arbitrary arrangement,
which may have implications for possible design of future devices
based on the manipulation of artificially pinned vortices. In this
report we present such an investigation facilitated by arranging the
holes in a line centred along the main axis of a superconducting
microbridge [lower inset to Fig.~1(a)], which eliminates the matching
phenomena but still provides a sufficient signal arising from the
presence of the holes. In addition, we have minimised the influence
of the intrinsic pinning by using the amorphous superconductor
Nb$_{0.7}$Ge$_{0.3}$, which is known for its very low background
pinning.\cite{basel,lopaper}

The vortex motion was detected by recording the electric field vs
current density characteristics $E(J)$ in three different temperature
regimes, over a wide range of applied currents, and $B$ up to the
upper critical magnetic field $B_{c2}$. Close to $T_c$ we 
observed a clear {\it increase of the power dissipation by the holes}
over the whole range of $B$ and $J$. This result can be explained
plausibly by taking into account a local enhancement of the vortex
driving force due to a spatial modulation of the driving-current
density around the holes. To our knowledge such current-modulation
effects have so far been disregarded in the interpretation of previous
experiments carried out on perforated superconductors.
 Well below $T_c$ the holes pin the vortices and reduce the
dissipation, which shows that their usefulness as APCs may extend
to low temperatures if the intrinsic pinning is weak. Our
results imply that the interplay of hole pinning and inhomogeneous
current drive determines whether the holes enhance or reduce the
dissipation in the vortex transport.

%
\begin{figure}
\includegraphics[width=80mm]{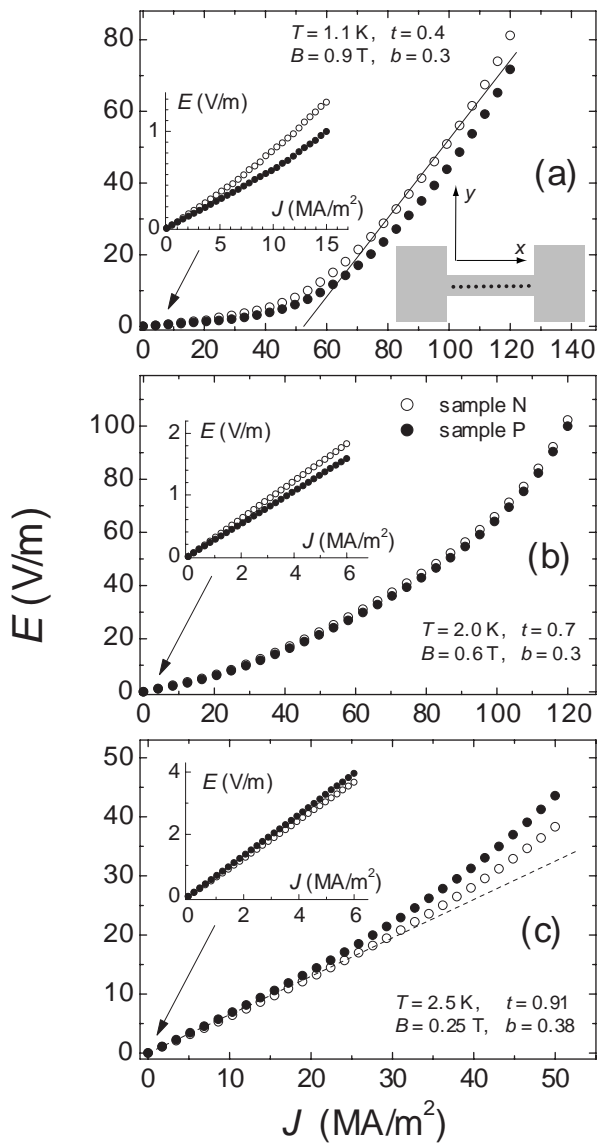}
\caption{$E(J)$ of sample N ($\circ$) and sample P ($\bullet$) for
(a) $T=1.1$ K , $B=0.9$ T, (b) $T=2.0$ K, $B=0.6$ T and (c) $T=2.5$ K, $B=0.25$ T. The
solid and dashed lines are plots of LO $\rho_f$ expected at low and high $t$, respectively.
Lower inset to (a): A sketch of sample P (not to scale) and the designation of the directions
(see the text). Upper inset to (a) and insets to (b), (c): The initial parts of the $E(J)$ in an
expanded scale for a better view.}
\end{figure}

Two 210 $\mu$m long, 5 $\mu$m wide and 20 nm thick microbridges between two large contact pads were
sputtered onto the same Si/SiO$_2$ substrate as described in
Ref.\onlinecite{basel}. Both samples had the same $T_c =$ 2.75 K and
the same overall normal-state and superconducting properties.
Throughout the paper we use the nonperforated sample (sample N) as a
reference for identifying the effects of holes. Its transport
properties were analysed in detail in Ref.\onlinecite{lopaper}, and
its GL parameters $\xi(0)= 6.8$ nm and $\lambda(0) = 1.15$ $\mu$m are
taken as representative of both samples. The perforated sample
(sample P) contained a line of equidistant holes along the central
axis of the microbridge (designated as $x$ direction) as sketched in
the lower inset to Fig.~1(a). The holes had a diameter of 800 nm
[slightly smaller than $\lambda(0)$] and their centre-to-centre
distance was 1.2 $\mu$m. A magnetic field was applied perpendicularly
to the film plane and a dc current was passed in the $x$ direction,
so that the vortices traversed the sample in the $y$ direction. The
current sweep rate was 10 nAs$^{-1}$. The applied current is for
sample P converted into the {\it average} current density $J_P$ by
calculating the (uniform) current density $J_N$ for sample N and
taking the ratio of the normal-state resistances to determine $J_P =
1.23 J_N$.  We use this notation, together with $E_N$, $E_P$ for the
electric field in samples N and P, respectively, when referring to the
$E(J)$ of the two samples specifically. At high $J$ the
$E(J)$ exhibit nonlinearities due to the flux-flow instabilities
that are related to the
nonequilibrium changes of vortex cores.\cite{lopaper} Here we
concentrate on the close-to-equilibrium regime where these effects
have not yet been developed and the vortex cores maintain their equilibrium
properties.

In Fig.~1 we plot typical $E(J)$ representative of three temperature
regimes: (a) $T$ well below $T_c$ (1.1 K, $t=0.4$), (b) intermediate
$t$ (2.0 K, $t = 0.7$), (c) $T$ close to $T_c$ (2.5 K, $t=0.91$). The
open circles show $E_N (J_N)$, solid circles $E_P (J_P)$, and the
lines different theoretical predictions of the Larkin-Ovchinnikov
(LO) flux flow (FF) theory,\cite{lo} as discussed below. The scaled
magnetic field $b = B/B_{c2}$ corresponds to 0.30 - 0.38, the
$B_{c2}$ values being  (a) 3.0 T, (b) 2.0 T and (c) 0.65 T, with the
uncertainty of around 5 \%. At low $T$ [Fig.~1(a)] the $E(J)$ of both
samples reveal two different dynamic regimes: thermally activated
magnetoresistance at $J \rightarrow 0$, followed at larger $J$ by an
$E \propto (J - J_c)$ behaviour that implies FF against a background
pinning potential with a depinning threshold $J_c$.\cite{basel} The
solid line is a plot of the latter dependence, with the slope $dE/dJ
= \rho_n b/0.9$ equal to the low-$t$ LO FF resistivity $\rho_f$
($\rho_n$ is the normal-state resistivity), and $J_c$ chosen to
obtain a fit to $E_N(J_N)$. The LO theory describes this part of
$E(J)$ reasonably well until the out-of-equilibrium nonlinearities in
$E(J)$ start to take place at large $J$.\cite{lopaper} As can be
seen, the power dissipation in sample {\rm P} is lower for $\sim 10$
\% throughout the whole current-density range, suggesting an efficient
pinning by the holes even far below $T_c$.  In Fig.~1(b) we show the
$E(J)$ at $T=2.0$ K ($t=0.7$), just at the boundary of the low-$t$
and high-$t$ regimes, displaying a weaker reduction of $E_P$ below
$E_N$. Close to $T_c$ [Fig.~1(c)] the theoretical high-$t$ LO
$\rho_f$ (see Ref.\onlinecite{lopaper} for details), shown by the
dashed line, describes $E_N(J_N)$ excellently starting from $J=0$ and
up to the appearance of the nonlinearities mentioned before. However,
over the entire current-density range {\it the dissipation in sample P is
larger than in sample N}, unexpectedly and in contrast to the result
at lower temperatures. The reason for this peculiar behaviour cannot
be found in the pinning properties of holes.

%
\begin{figure}
\includegraphics[width=65mm]{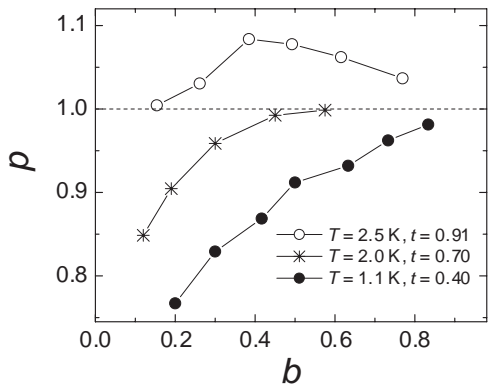}
\caption{The ratio $p = P_P/P_N$ of the power dissipated in samples
$P$ and $N$, integrated up to the appearance of the flux flow
instabilities, vs $b=B/B_{c2}$.
Sufficiently below $T_c$ the holes reduce the dissipation ($p<1$)
whereas close to $T_c$ the situation is opposite.}
\end{figure}

The magnetic field range over which $E_N \neq E_P$ is wide at all
three characteristic temperatures. This is demonstrated in Fig.~2,
where we plot the ratio $p = P_P / P_N$ of the power density
dissipated in samples P and N vs
$b=B/B_{c2}$. The integration $P_{N,P} = \int E_{N,P} dJ_{N,P}$ is
for each curve performed over the maximum region where
the dissipation is not affected by
the FF instabilities.
We have checked whether $p$ depends on the upper limit of
integration, and we found minor numerical differences of the order of
10 - 15 \% of the values shown in Fig.~2, with no change in the
general shape of $p(b)$. As can be anticipated from the $E(J)$
shown in Figs.~1(a),1(b), for $t=0.4$ and $t=0.7$ we find $p < 1$, i.e.
the holes are active
pinning sites and decrease the dissipation. Their relative
contribution to the pinning becomes smaller as $t$ and $b$
increase, which is expected qualitatively and discussed later. Close
to $T_c$ the dissipation in sample P is always larger, thus $p
>1$ with a maximum around $b\sim 0.4$. This enhancement implies
a suppression of the artificial pinning by another effect which is
fully manifested close to $T_c$.


%
\begin{figure}
\includegraphics[width=70mm]{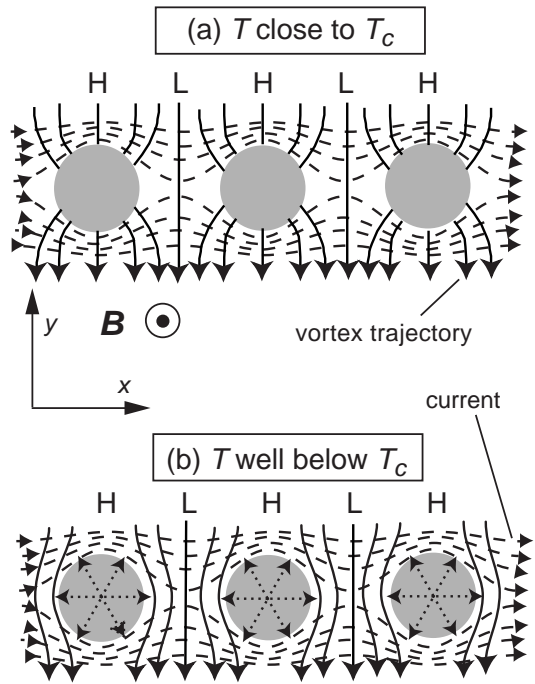}
\caption{A qualitative consideration of the vortex trajectories
(arrowed solid lines) around the holes at (a) $T$ close to $T_c$ and
(b) $T$ well below $T_c$. The modulated current is indicated by the
arrowed dashed lines, the density of which represents the magnitude
of $J$. Arrowed dotted lines in (b) depict the repulsion excerted by
the vortices pinned by the holes to other vortices. The labels L
and H denote regions of low and high vortex flow density,
respectively.}
\end{figure}

In order to understand the above results one has to address the
equation of motion for the vortices
\begin{equation}
\label{eqmotion}
m \dot{{\bf u}} = {\bf F}_d - \eta {\bf u} + {\bf F}_h  \; \; \; ,
\end{equation}
where ${\bf u}$ is the vortex velocity, $\eta$ the vortex-motion
viscosity, $m$ the effective vortex mass per unit length,\cite{mass}
${\bf F}_d = \phi_0 {\bf J} \times  {\bf \hat{z}}$ the driving force
(${\bf \hat{z}}$ is the unit vector in the direction of the applied
magnetic field), and ${\bf F}_h$ being the repulsive force between
the mobile vortices and those pinned by the holes. The spatial
dependence of the {\it i}-th component of the vortex acceleration
$\dot{{\bf u}}$ is given by $\dot{u}_i = {\bf u} \cdot \nabla u_i$.
For our qualitative reasoning we have neglected other contributions
(e.g. the interaction of mobile vortices with each other or with
intrinsic pinning potential). Knowing the supercurrent distribution
${\bf J}({\bf r})$ one can solve Eq.~(\ref{eqmotion}) to find the
vortex-velocity profile ${\bf u}({\bf r})$. The generated electric
field, the $x$ component of which contributes to measured $E_P$, is
calculated as ${\bf E}({\bf r}) = {\bf u}({\bf r}) \times {\bf
B}({\bf r})$, where ${\bf B}({\bf r})$ is related to the local vortex
density $n({\bf r}) = B({\bf r})/ \phi_0$. When confining a pinned
vortex each hole acts as a source of repulsion to other vortices,
with ${\bf F}_h$ pointing radially from the centre of the hole. The
repulsive force between two vortices separated by $r \ll \lambda$ is
given by $\lambda^{-2} \ln (\lambda /r)$ and therefore weakens as
$\lambda \rightarrow \infty$. Furthermore, the probability of a
vortex being pinned by a hole becomes progressively smaller as
$\lambda$ grows much larger than the hole diameter.\cite{take} Hence
it is reasonable to assume that ${\bf F}_h$ decreases monotonically
with increasing $\lambda$ and eventually becomes irrelevant in the
limit of diverging $\lambda$ at $T \rightarrow T_c$. In the following
we discuss the results for $t=0.91$ assuming ${\bf F}_h = 0$.

We note that the driving force ${\bf F}_d$ is {\it not uniform around
the holes} because the supercurrent density $\bf{J}$ is spatially
modulated. A schematic of the modulation of $\bf{J}$, and
consequently of ${\bf F}_d$, around the holes is shown in Figs.~3(a)
and 3(b) by the dashed lines. In the regions denoted by H the
driving force is larger than far from the holes, while in the regions
denoted by L the situation is the reverse. With ${\bf F}_h \approx
0$ in the vicinity of $T_c$, the vortices are accelerated parallel to
${\bf F}_d$ (perpendicular to ${\bf J}$), thus the vortex
trajectories accumulate in regions H as sketched in Fig.~3(a) by the
solid lines. The second effect of the current modulation is that in
regions H the vortices move faster than in the sample bulk and produce a
larger local electric field, while in regions L the opposite happens.
Although the holes themselves shorten the distance the fast vortices
travel in regions H, a sufficient imbalance in favour of the number
of these vortices, together with their large acceleration, may result
in the total dissipation in sample P being larger than in sample N.
Therefore if a highly dense two-dimensional array of holes\cite{welp}
is intended to be used for enhancing the pinning of a sample there is
no guarantee that this will work close to $T_c$ in the presence of a
transport current, although the magnetisation loops may widen up.
Moreover, the current-induced vortex-velocity enhancement strengthens
with increasing $J$, which sheds more light on the similar result of
Ref.\onlinecite{jiang} which also occurred at a relatively large
applied current.

The nonmonotonic $b$ dependence of $p > 1$ may be linked to the
competition between the enhanced current drive around the holes and
the mutual repulsion of the mobile vortices, as explained below. The
excess electric field relative to that far from the holes is
estimated as $\Delta E \sim \phi_0 \Delta u \Delta n$, where $\Delta
u$ and $\Delta n$ are the effective excess vortex velocity and
density in regions H, respectively. While $\Delta u$ depends on ${\bf J}$,
$\eta$, $m$, and is independent of $B$, $\Delta n$
also depends on the repulsive interaction between the mobile vortices,
which is weak at low $b$ and strong at high $b$.
At low fraction $b=B/B_{c2}$
of the volume filled with the vortices they can all easily be
channeled through regions H. Thus a larger vortex density results in
the larger $\Delta n$ and $\Delta E$. On the other hand, $\Delta n$
is at high $b$ limited by the reduced space available for the
channeling, which decreases $\Delta E$ and in turn results in a
nonmonotonic $p(b)$.

As temperature is lowered  $\lambda$ becomes smaller and the
holes start to pin more efficiently,\cite{take} which explains
qualitatively the stronger reduction of the dissipation at lower
temperatures (Fig.~2). The vortices trapped by the holes
now repel the incoming vortices and change the vortex trajectories
depicted in Fig.~3(a) to those shown in Fig.~3(b). The repulsive
force ${\bf F}_h$ is indicated by the dotted arrows. In a simple
model of pinning by the holes in dynamic conditions, a vortex remains
pinned by a hole for some time until it is replaced by an incoming
vortex. At low vortex density almost all vortices in the
vicinity of holes are either pinned by them or scattered to pass
through the regions L of low current density. Hence, the holes cause
a noticeable reduction of the dissipation. As $B$ increases the
incoming vortices exert more force upon those pinned at the holes,
reduce the pinning time, and the overall suppression of the
dissipation decreases. This qualitative picture may explain the
behaviour of $p(b)$ at low temperatures.


In conclusion, we have patterned a line of holes with a diameter
close to $\lambda(0)$ along an amorphous Nb$_{0.7}$Ge$_{0.3}$
microbridge. A comparison of the measured $E(J)$ curves of the
samples with and without perforation reveals an unusual crossover in
the power dissipation close to and well below $T_c$. Close to $T_c$
the artificial pinning is weak and an unexpected rise of the
dissipation is observed. This is attributed to the inhomogeneous
current distribution around the holes leading to a significant
increase of the local vortex velocity. As temperature is lowered the
pinning by the holes becomes stronger and eventually suppresses the
vortex velocity enhancement. In addition, our weak background pinning
has permitted, to our knowledge, the first observation the pinning
properties of holes as artificial pinning centres far below $T_c$.

We thank B.~Stojetz,  A.~Bauer, F.~Rohlfing, W.~Meindl and
M.~Furthmeier for their technical assistance. This work was partly funded by the
Deutsche Forschungsgemeinschaft within the Graduiertenkolleg 638. We
gratefully acknowledge additional support by the Croatian Ministry of
Science (Project No. 119262) and the Bavarian Ministry for Science,
Research and Art.

\end{document}